\documentstyle[preprint,aps]{revtex}
\begin{document}
\draft
\tightenlines
\preprint{IP/BBSR/95-29}
\title{
\bf{ Mirror Symmetry in the Few Anyon Spectra in a Harmonic
Oscillator Potential
} }
\author{ Ranjan Kumar Ghosh}
\address{ Acharya Brajendra Nath Seal College,
Cooch Behar, 736 104 India}

\author{ Sumathi Rao}
\address{ Institute of Physics,
Bhubaneswar 751 005, India}
\maketitle
\begin{abstract}
We find that the energy spectra of four and five anyons in a
harmonic potential exhibit some mirror symmetric (reflection
symmetric about the semionic statistics point $\theta=\pi/2$)
features analogous to the mirror symmetry in the two and three
anyon spectra. However, since the $\ell=0$ sector remains
non-mirror symmetric, the fourth and fifth virial coeffients do
not reflect this symmetry.
\end{abstract}
\pacs{03.65.Ge,~~05.30.-d,~~11.30.Pb}


\def\q4i{Q_4^{i}}
\def\d{\Delta}
\def\modd{\vert \Delta \vert}

\def\ddz{{\partial\over \partial z_n}}
\def\ddzs{{\partial\over \partial z_n^*}}
\def\ddzm{{\partial\over \partial z_m}}
\def\ddzms{{\partial\over \partial z_m^*}}
\def\ddu{{\partial\over \partial u_i}}
\def\ddus{{\partial\over \partial u_i^*}}

\def\zm{z_m}
\def\zms{z_m^*}
\def\ui{u_i}
\def\uis{u_i^*}
\def\vi{v_i}
\def\vis{v_i^*}
\def\us{u_1^*}

\def\o{\omega}
\def\eps{\epsilon}
\def\al{\alpha}
\def\alim{\alpha_{im}}
\def\epsf{\eps_{ijkl}}
\def\ai{a_i}
\def\bi{b_i}
\def\aid{a_i^{\dagger}}
\def\bid{b_i^{\dagger}}
\def\ad{a_1^{\dagger}}
\def\bd{b_1^{\dagger}}

\def\and{a_n^{\dagger}}
\def\amd{a_m^{\dagger}}
\def\bod{b_o^{\dagger}}

\def\aj{a_j}
\def\bj{b_j}
\def\ajd{a_j^{\dagger}}
\def\bjd{b_j^{\dagger}}

\def\am{a_m}
\def\bm{b_m}
\def\amd{a_m^{\dagger}}
\def\bmd{b_m^{\dagger}}

\def\ak{a_k}
\def\bk{b_k}
\def\akd{a_k^{\dagger}}
\def\bkd{b_k^{\dagger}}

\def\sm1{\sum_{i=1}^{N-1}}
\def\be{\begin{equation}}
\def\ee{\end{equation}}
\def\bea{\begin{eqnarray}}
\def\eea{\end{eqnarray}}

The multi-anyon problem continues to remain interesting as it
defies both analytic and numerical attempts to solve it. As has
been often reiterated in the literature \cite{anyorig,anyrevs},
the difficulty stems from the inability to write multi-anyon
wave-functions as simple products of single anyon
wave-functions. Hence, even simple quantum mechanical problems
involving more than two anyons remain insoluble.

The two standard routes that have been followed to tackle the
problem have been either to start from the high density limit or
from the low density limit. In the high density limit
\cite{anysup}, the anyons are stripped of their fluxes, which
get spread out and the mean field problem involves fermions or
bosons moving in a uniform magnetic field. This is the approach
that led to anyon superconductivity. Standard
improvements in the mean field theory by incorporation
of higher order corrections have been made; however, in the
absence of a small expansion parameter, the validity of these
approximations remains unclear.

In the low density limit \cite{anystat}, the strategy has been
to solve few anyon problems.  For more than two anyons, the
problem cannot be solved exactly. However, for free anyons,
anyons in a magnetic field and anyons in a harmonic potential,
several exact wave-functions of the Hamiltonian can be
constructed\cite{wavefn}. But since these wave-functions do not
form a complete set, they cannot be used to calculate
statistical quantities like the virial coefficients. For this,
perturbation theory \cite{anypert} about the bosonic or
fermionic limit has been tried using the statistical parameter
as the expansion parameter. Also, the lowest lying eigenstates
of the Hamiltonian have been obtained numerically for the three
\cite{3num} and four \cite{4num} anyon problems. These
approaches gave some insight into the virial coefficients and
hence the statistical mechanics of a gas of anyons.

More recently, a useful approach in the few anyon system was
pioneered by Sen \cite{diptiman}, who showed that by studying
the symmetries of the Hamiltonian of the few anyon system, some
exact statements could be proven about the spectrum. He showed
that the three anyon spectrum in a harmonic potential is
(almost) mirror symmetric about the semionic point
($\theta=\pi/2$) by constructing a fermionic operator $Q_3$ (a
quadratic polynomial) that commuted with the Hamiltonian. More
interestingly, by using this symmetry, he was able to show that
the third virial coefficient is mirror-symmetric about
$\theta=\pi/2$. However, so far, no exact statements have been
made for more than three anyons.

With the view that any exact statement that can be made about
the many anyon spectrum is a useful excercise, in this paper,
we study four and five anyon spectra in a harmonic potential in
some detail. In analogy with Ref.\cite{diptiman}, we construct
fermionic operators. For the four anyon case, we construct four
fermionic operators $Q_4^{i}$, which are cubic polynomials.
These operators do not commute with the Hamiltonian - rather,
they act as raising and lowering operators. We show that all
states (other than the $\ell=0$ states and those linear states that
are annihilated by $Q_4^{i}$) have a `skewed' mirror symmetry
about $\theta=\pi/2$ - $i.e.$, a state with energy $E$ and
angular momentum $\ell$ is paired by the $\q4i$ with states
with energy $E \pm \omega$ and angular momentum $\ell \pm 1$. For
the five anyon case, however, a unique fermionic operator $Q_5$
can be constructed, (a fourth order polynomial), which commutes
with the Hamiltonian and whose action can be used to demonstrate
a genuine mirror symmetry of the spectrum for all states (other
than $\ell=0$ states and those linear states annihilated by $Q_5$).
However, the absence of the mirror symmetry for the $\ell=0$ states
(which includes non-exactly solved states) implies that this is
not sufficient to make any exact statement about the
fifth virial coefficient.

We start by considering the problem of $N$ anyons in a harmonic
potential. This is a generic problem, since results for anyons
in a magnetic field as well as free anyons can be derived from
these results. We shall work in the bosonic gauge where the
wave-function is bosonic and the information about the
fractional statistics is incorporated in the Hamiltonian. The
transformation to the anyon gauge is simply achieved by changing
the Hamiltonian to a free Hamiltonian and multiplying the
wave-functions by the phase $\d/\modd$ where
\begin{equation}
\d = \Pi_{n<m}^N (z_n - z_m)
\end{equation}
and $z_n=(x_n+i y_n)/\sqrt 2$ are complex coordinates on a
plane. The Hamiltonian in the bosonic gauge is given by
\begin{equation}
{\cal H} = -\sum_{n=1}^N [ (\ddz + s_n) (\ddzs - s_n^*) +\omega^2 z_n
z_n^*]
\end{equation}
where we have set the mass of the particles $m=1$ and the gauge
potentials are given by
\begin{equation}
s_n = {\alpha\over 2} \sum_{n \ne m}^N {1\over z_n-z_m} =
{\alpha\over 2} \ddz {\rm log} \d.
\end{equation}
Here, $\alpha$ is the statistics parameter , ( = $\theta/\pi$)
and ranges from $\alpha=0$ to $\alpha=1$.
Since the gauge potentials are functions only of the relative
coordinates, it is more convenient to use Jacobi coordinates
defined by
\begin{eqnarray}
u_0 & = & {1\over \sqrt N} \sum_{n=1}^N z_n \equiv \alpha_{0m}
z_m, \nonumber \\
u_i & = & {1\over \sqrt {i(i+1)}} (z_1+z_2 + \cdots -i ~~z_{i+1})
\equiv \alpha_{im} z_m
\label{eq:jacobi}
\end{eqnarray}
with $i=1,\cdots,N-1$. In terms of these coordinates, the
Hamiltonian neatly splits into a CM part $H_{CM}$ which is a
free oscillator
and a relative part $H$ which is given by
\begin{equation}
H= -\sum_{i=1}^{N-1} [(\ddu + v_i) (\ddus - v_i^*) + \omega^2 u_i
u_i^*]
\end{equation}
with the gauge potentials $v_i$ being given by
\begin{equation}
v_i = {\alpha\over 2} \ddu {\rm log} D, i=1,\cdots, N-1
\end{equation}
and $D\equiv D(u_i)$ is the transformed form of $\d$.
This is easily proved using the
orthogonality of the transformation in Eq. \ref{eq:jacobi} and
the analyticity of $s_n$ and $v_i$.
Note that $D$ is independent of $u_0$, since it only depends on
relative coordinates.  (A slightly different form
of $v_i$ is used in Ref.\cite{diptiman}.)

\newpage

Let us now introduce the ladder operators $a_i, a_i^{\dagger},
\bi$ and $\bid$ -
\begin{eqnarray}
\ai & = &{1 \over \sqrt 2} (\ddu + v_i + \omega \uis), \quad
\bi = {1\over \sqrt 2} (-\ddus + v_i^* - \omega \ui), \nonumber \\
\aid & = & {1\over\sqrt 2} (-\ddus + \vis + \omega \ui), \quad
\bid = {1\over\sqrt 2} (\ddus + \vis - \omega \uis),
\end{eqnarray}
for $i=1,\cdots, N-1$. The $\ai$ and $\aid$ commute with $\bi$
and $\bid$ and
\be
[\ai,a_j^{\dagger}] = \o \delta_{ij} , \quad [\bi,
b_j^{\dagger}] = \o \delta_{ij}.
\ee
The Hamiltonian and the relative angular momentum operators $L
= \ui \ddu - \uis \ddus$ can be written in terms of these
operators as
\begin{eqnarray}
H & = &  \sm1 (\aid \ai + \bid \bi) +(N-1) \o \nonumber \\
{\rm and} \quad
\o L & = & \sm1 (\aid \ai - \bid \bi) - {N(N-1) \over 2} \alpha.
\label{eq:handl}
\end{eqnarray}
Although the Hamiltonian may look trivially solvable, note that
the ladder operators are not the conventional ones. Firstly,
they transform as an $N-1$-dimensional irreducible
representation (IR) of the permutation group $S_N$ - $i.e.$, if
$|\psi\rangle$ is a bosonic state, then $\aid |\psi\rangle$ or
$\bid |\psi\rangle $
are not bosonic or fermionic , but transform as an
$N-1$-dimensional IR of $S_N$.  Also, the wavefunctions of the
Hamiltonian must vanish as $|z_i - z_j|^{\alpha}$ or faster as
$z_i$ approaches $z_j$. This is the hard-core constraint on the
bosons which can be deduced from the statistics of the anyons in
the anyon gauge. Such wave-functions are called physical or
non-singular. But there is no guarantee that $\aid$ and $\bid$
acting on physical wavefunctions will only lead to physical
wavefunctions. They could lead to singular wave-functions which
must be rejected.

However, Sen~\cite{diptiman} showed that polynomials of these
ladder operators could be formed, which transform as bosons
($Q_B$) or fermions ($Q_F$). He showed that of ten possible
quadratic bosonic operators, $K_+ = 2 \aid \bid$ and $K_- = 2
\ai \bi$ are `good' operators - $i.e.$, they do not produce
unphysical states when acting on physical states - and along
with $K_3 = H$, they form an $SO(2,1)$ algebra \cite{so21} given
by
\be
[H,K_{\pm}] = \pm 2 \o K_{\pm}.
\ee
This shows that the spectrum gets organised in terms of
$SO(2,1)$ families, with members within each family differing
in energy from each other by $2 \o$.

For the three anyon problem, he was also able to prove a much
more interesting result by constructing a `good' fermionic
operator
\be
{\widetilde Q_3} = a_1^{\dagger} b_2 - a_2^{\dagger} b_1.
\ee
Since this operator is fermionic, it turns bosons into fermions
and vice-versa - $i.e.$, it changes the statistics parameter
$\alpha$ to $\alpha+1$. A parity operator $P$ can also be
defined which maps $1+\alpha$ to $1-\alpha$ and hence the
combined operator $ Q_3 = {\widetilde Q_3} P$ maps bosonic
states at statistics parameter $\alpha$ to bosonic states at
statistics $1-\alpha$, (which are equivalent to fermionic states
at statistics $-\alpha$). Since, this operator commuted with the
Hamiltonian, he essentially proved that the three anyon spectrum
is mirror symmetric about the semionic point $\alpha = 1/2$
(except for some of the linear exactly solved states, which were
annihilated by $Q_3$). He further used this symmetry
to prove that the third virial coefficient is exactly mirror
symmetric.

We now try to extend these results to the case when $N=4$. Our
aim is to construct fermionic or bosonic operators that enable
us to make some exact statements about the spectrum. Bosonic
bilinears (quadratic operators) can be constructed as before and
the $SO(2,1)$ symmetry can be extended to the $N=4$ case.
But unlike the three anyon case, no fermionic
bilinear operator can be constructed. However, consider the
following trilinear operators -
\begin{eqnarray}
Q_4^1 & = & \eps_{ijk} \aid \bj \bkd \nonumber \\
Q_4^2 & = & \eps_{ijk} \ai \ajd \bkd \nonumber \\
Q_4^3 & = & \eps_{ijk} \ai \ajd \bk  \nonumber \\
Q_4^4 & = & \eps_{ijk} \ai \bj \bkd. \label{eq:q4}
\end{eqnarray}
By construction, these operators are anti-symmetric under $u_i
\leftrightarrow u_j$. But they can also be proven to be
antisymmetric under $z_m \leftrightarrow z_n$ in the following
way. As far as the orthogonal transformation in
Eq.\ref{eq:jacobi} is concerned, the coordinates $u_i$ and
$u_i^*$ and the derivatives $\ddu$ and $\ddus$ transform with
the same coefficients, -$i.e.$,
\bea
\ui &\rightarrow & \alim z_m \nonumber \\
\uis &\rightarrow & \alim \zms \nonumber \\
\ddu &\rightarrow & \alim \ddzm \nonumber \\
\ddus &\rightarrow & \alim \ddzms,
\eea
where $\alim$ are defined in Eq.\ref{eq:jacobi}. Moreover,
the $v_i$ and $\vis$ also transform with the same coefficients
because
\bea
v_i &=& {\al\over 2} \ddu {\rm log} D = {\al\over 2 D} \ddu D
\rightarrow {\al\over 2\Delta} \alim \ddzm \Delta \nonumber \\
{\rm and} \quad
\vis &=& {\al\over 2} \ddus {\rm log} D^* = {\al\over {2 D^*}}
\ddus D^* \rightarrow {\al\over 2{\Delta^*}} \alim \ddzms \Delta^* .
\eea
This implies that we can now have new oscillators $a_m$ and
$b_m$ defined by
\bea
\ai &=& \alim \am \nonumber \\
\bi &=& \alim \bm \nonumber \\
\aid &=& \alim \amd \nonumber \\
\bid &=& \alim \bmd,
\eea
where $m = 1,..4$. In terms of these oscillators, the $Q_4^i$
can be written as
\be
Q_4^i = \eps_{jkl} \alpha_{jm} \alpha_{kn} \alpha_{lo} O_4^i =
\left\vert \begin{array} {ccc}
\al_{1m} ~~~\al_{2m} ~~~\al_{3m} \\
\al_{1n} ~~~\al_{2n} ~~~\al_{3n} \\
\al_{1o} ~~~\al_{2o} ~~~\al_{3o}
\end{array} \right\vert O_4^i
\equiv \beta_{mno} O_4^i \label{eq:det}
\ee
where $O_4^i = \amd b_n \bod, a_m \and \bod, a_m \and b_o$ and
$a_n b_m \bod$ respectively for $i=1,\cdots 4$. Using the
coefficients defined in Eq.\ref{eq:jacobi}, it is easy to
check that the determinant $\beta_{mno}$ is completely
antisymmetric in $m,n$ and $o$ and in fact, can be written as
$\beta_{mno} = \epsilon_{mno}/2$. This proves that the operators
$Q_4^i$ are completely antisymmetric under $z_m\leftrightarrow
z_n$. (A similar proof can be constructed for the antisymmetry
of $Q_3$ using $\epsilon_{ij} \alim \al_{jn} = \al_{mi}^{T}
\epsilon_{ij} \al_{jn} = \epsilon_{mn}/ {\sqrt 3}$.  This is
somewhat different from the proof constructed by
Sen\cite{diptiman}.)
Furthermore, these operators act as raising and lowering
operators of the Hamiltonian and angular momentum -
\begin{eqnarray}
\left[H, Q_4^i\right] & = & \o Q_4^i, \quad i=1,2 \nonumber \\
\left[H, Q_4^i\right] & = & - \o Q_4^i, \quad i=3,4 \nonumber \\
\left[L, Q_4^i\right] & = & Q_4^i, \quad , i=1,3 \nonumber \\
\left[L, Q_4^i\right] & = & - Q_4^i, \quad, i=2,4.
\end{eqnarray}

Are these `good' operators? To answer this question, notice that
we only have to examine the singularity as $z_1
\leftrightarrow z_2$, ($i.e.$, $u_1 \rightarrow 0$), since
the operators have already been shown to be antisymmetric.
As $u_1 \rightarrow 0$, in any anyonic theory, physical
wavefunctions can vanish as
\begin{eqnarray}
i&)& ~~(u_1 \us)^{|\al|/2}, \quad  \ell = 0 \nonumber \\
ii&)& ~~(u_1 \us)^{\al/2} u_1^{\ell} \quad \ell  \ge  2 \quad
(\ell={\rm even}) \nonumber \\
iii&)& ~~(u_1 \us)^{-\al/2} u_1^{*\ell} \quad \ell  \ge  2 \quad
(\ell={\rm even}) \nonumber \\
iv&)& ~~(u_1 \us)^{\al/2} u_1^{\ell} ~+~ (u_1
\us)^{-\al/2} u_1^{*\ell}\quad \ell  \ge  2 \quad
(\ell={\rm even}),
\end{eqnarray}
where, in case iv), we mean that the two terms
can be multiplied by different non-vanishing functions of the
remaining $\ui$. For $0<\al<1$, it is easy to see that $\ad$
and $b_1 ( \simeq {1\over\sqrt 2} ({\partial\over \partial
\us} - v_1^*)$ when $u_1\rightarrow 0$) never produce any
singularities when acting on any of the wave-functions (i)
through (iv). But single powers of $a_1$ and $\bd$ produce
singular wave-functions when they act on the $\ell=0$
wave-functions in (i). Since, all the $Q_4^i$ contain either the
term $a_1$ or $\bd$, they are not `good' operators per se;
however, they produce singularities only when they act on the
$\ell = 0$ wave-functions.

We can also find the states on which the action of the
operators $Q_4^i$ gives zero as follows.
Any state annihilated by $a_i$ will be annihilated by the
sum $2 \aid a_i$, which in turn, (from Eq.\ref{eq:handl})
implies that its energy will be given by
\be
E = (N-1) \omega - (L+{\al\over 2} N(N-1)) \omega,
\label{eq:e1}
\ee
- $i.e.$, they are states with energies linearly falling with
$\al$.  Similarly, any state annihilated by $b_i$ will have an
energy dependence that linearly rises with $\al$ -
\be
E = (N-1) \omega + (L+{\al\over 2} N(N-1)) \omega.
\label{eq:e2}
\ee
Thus $Q_4^1$ annihilates states with linearly rising energies,
$Q_4^2$ annihilates states with linearly falling energies and
$Q_4^3$ and $Q_4^4$ annihilate states with both linearly rising
and falling energies.
Moreover, we can also check that if $Q_4^i|\psi\rangle$
vanishes, then $[Q_4^i, K_+] |\psi\rangle$ also vanishes.
Hence, the entire $S0(2,1)$ family obtained by acting
$K_+$ on the appropriate linear state or states is annihilated by
$Q_4^i$. But unlike the three anyon case, here
all the states annihilated by $Q_4^i$ need not be of
the above form -$i.e.$, need not be linear states.

What does all this tell us about the spectrum ? For all states
at statistics parameter $\al$
other than the linear states (and perhaps some
non-linear states) that are
annihilated by $Q_4^i$ and the $\ell=0$ states, there exist
partner states at statistics $1-\al$ with energies $E \pm \o$
and angular momentum $\ell \pm 1$. Hence, given the spectrum at
$\al$, we can (almost) predict the spectrum at $1-\al$. This is
what we mean by saying that the problem of four anyons in a
harmonic potential has a `skewed' mirror symmetry. (This is
similar to the symmetry found in the two anyon
case\cite{chitra}.)

The more interesting case occurs when $N = 5$. Here, we can
construct a unique operator
\be
Q_5 = \epsf \aid \aj \bkd b_l \label{eq:q5}
\ee
which can be proven to be anti-symmetric under $z_m
\leftrightarrow z_n$ by a straightforward extension of the
argument that we used for the four anyon case. The determinant
in Eq.\ref{eq:det} is replaced by a $4\times 4$ determinant
$\beta_{mnop}$,which can be shown to be equal to
$\eps_{mnop}/{\sqrt 5}$, using the Jacobi coefficients in
Eq.\ref{eq:jacobi}.
Moreover, it commutes with both the Hamiltonian and the angular
momentum -
\be
[H,Q_5] = 0 \quad {\rm and} \quad [L,Q_5] = 0.
\ee
However, just as the $Q_4^{i}$, it is not a `good' operator per
se, because some terms in $Q_5$ give rise to singularities as
$u_1\rightarrow 0$, when acting on the $\ell=0$ wavefunctions in
$(i)$. $Q_5$ is non-singular when acting on all other physical
wavefunctions. Also, just as for the $Q_4^{i}$, we can show
that $Q_5$ annihilates linear states, - states with energies of
the form given in Eqs.\ref{eq:e1} and \ref{eq:e2}  because
each term in $Q_5$ contains both $\ai$ and $\bi$ for some $i$.
It is also easy to
check that $[Q_5,K_+]=0$. Hence, all members of the $SO(2,1)$
family formed from the base linear states
are annihilated by $Q_5$.

Hence, we conclude that {\it all} states except the $\ell=0$
states and the linear states (and perhaps a few other non-linear
states annihilated by $Q_5$) of the five anyon
system come in pairs. Each state with energy $E$ and angular
momentum $\ell$ at a statistics parameter $\al$ is accompanied
by a state with the same $E$ and same $\ell$ at statistics
parameter $1-\al$. {\it In fact, we make the specific prediction
that if the five anyon spectrum were computed and all the
$\ell\ne 0$ non-linear states were plotted as a
function of the statistics parameter $\al$, then there would be
an exact mirror symmetry about $\al=1/2$.} Since, the $\ell \ne
0$ set of states is much larger than the $\ell=0$ set of
states, it would not be surprising if the fifth virial
coefficient shows a mirror symmetry about $\al=1/2$. However, to
really demonstrate that, we need to show that the linear
states (and any other states that may be annihilated by $Q_5$)
and $\ell=0$ states do not contribute to the difference in the
virial coefficients $a_5(\al) - a_5(1-\al)$. We have been unable
to show this so far.

We note that this generalisation of $Q_3$ cannot be extended to
cases beyond $N=5$, (four relative coordinates) because
there are only four types of oscillators, $\ai, \bi,\aid$ and
$\bid$, and no further antisymmetrisation is possible. It may
still be possible that for $N=7,9, \cdots$, the spectrum remains
almost mirror symmetric, but other methods will have to be found
to prove it.

In conclusion, in this paper, we have studied the symmetries of
the few-anyon spectrum in some detail. We have constructed new
fermionic operators (see Eqs.\ref{eq:q4} and \ref{eq:q5}). For
the four anyon problem, these operators act as raising and
lowering operators of the Hamiltonian, but for the five anyon
case, this fermionic operator commutes
with the Hamiltonian. We have been able to show for the five
anyon case that the system exhibits a mirror symmetry about
$\al=1/2$ (except for $\ell=0$ and linear states).
Thus, we have strengthened the conjecture\cite{diptiman} that
all odd virial coefficients are mirror symmetric.

\section*{Acknowledgments}
We thank Diptiman Sen for useful communications.
One of us (RKG) thanks the Institute of Physics, Bhubaneswar,
where this work was initiated, for hospitality.

\end{document}